# Stable multicolor periodic-wave arrays

Yaroslav V. Kartashov,[1,2] Alexey A. Egorov,[2] Anna S. Zelenina,[2] Victor A. Vysloukh,[3] Lluis Torner[1]

*[1]ICFO-Institut de Ciencies Fotoniques, and Department of Signal Theory and Communications, Universitat Politecnica de Catalunya, E-08034 Barcelona, Spain*
*[2]Physics Department, M. V. Lomonosov Moscow State University, 119899, Moscow, Russia*
*[3]Departamento de Fisica y Matematicas, Universidad de las Americas - Puebla, Sta. Catarina Martir, 72820, Puebla, Cholula, Mexico*

We study the existence and stability of cnoidal periodic wave arrays propagating in uniform quadratic nonlinear media and discover that they become completely stable above a threshold light intensity. To the best of our knowledge, this is the first example in physics of completely stable periodic wave patterns propagating in conservative uniform media supporting bright solitons.

Multiple-peaked periodic wave structures play a central role in Nature [1-6]. They are at the core of the development of modulational instabilities and turbulences in nonlinear systems, the eigenmodes of mechanical, molecular and electrical chains, Bloch waves in solid-state physics, the formation of matrices of ultracold atoms or trapped Bose-Einstein condensates, or the creation of light-induced reconfigurable photonic lattices. For example, self-sustained periodic wave structures find applications in Langmuir plasma waves [5,7], deep-water gravity waves [4,8], pulse trains in optical fibers [1-3,9], reconfigurable beam arrays in photorefractive crystals [6,10], matter waves in trapped Bose-Einstein condensates [11], or synchronously pumped optical parametric oscillators [12], to mention a few. By and large, the concept of periodic nonlinear waves allows to bridge the gap between localized solitons and linear harmonic or continuous waves [1,2].

However, to date *stable* periodic patterns are only known to occur in physical settings modeled by self-defocusing nonlinearities [3-5,7,8,13], in systems trapped by external potentials, like Bose-Einstein condensates in periodical traps [11], or in dissipative systems, like optical cavities. To the best of our knowledge, all previous examples of periodic (*or cnoidal*) waves in conservative uniform medium with self-focusing nonlinearity supporting bright solitons are dynamically unstable. Therefore, albeit some strategies to reduce the instability strength have been put forward recently, no examples of completely stable cnoidal waves are known in self-focusing media.

In this Letter we report the existence of the first known example of such *stable cnoidal wave families*. We discovered them in quadratic nonlinear media. Thus, they are *multicolored* and sustained by nonlinear cross-phase-modulations induced in parametric wave-mixing processes. Similar to the cnoidal waves existing, e.g., in self-focusing cubic nonlinear media, many cnoidal waves supported by quadratic nonlinearities suffer from modulational instabilities and thus self-destroy. However, we found that above a threshold field intensity the growth rate of the instabilities that affect so-called *cn*-type cnoidal waves suddenly vanishes completely, thus turning the cn-waves totally stable. Such unconditional stability was revealed by a mathematical formalism put forward to elucidate the linear stability of the cnoidal wave solutions, and was confirmed by direct numerical simulations in the presence of multiplicative broadband random initial perturbations.

Here we address the case of spatial light trapping in one-dimensional phase-matched second-harmonic generation, where a fundamental frequency wave (FF) and its second-harmonic (SH) interact with each other, but the concept is expected to hold in more general physical geometries and settings. The propagation of the corresponding slowly-varying envelopes under conditions for non-critical type I phase-matching is described by the system of reduced equations [14-16]

$$\begin{aligned} i\frac{\partial q_1}{\partial \xi} &= \frac{d_1}{2}\frac{\partial^2 q_1}{\partial \eta^2} - q_1^* q_2 \exp(-i\beta\xi), \\ i\frac{\partial q_2}{\partial \xi} &= \frac{d_2}{2}\frac{\partial^2 q_2}{\partial \eta^2} - q_1^2 \exp(i\beta\xi). \end{aligned} \tag{1}$$

Here $q_1 = (2k_1/k_2)^{1/2}(2\pi\omega_0^2\chi^{(2)}r_0^2/c^2)A_1$, $q_2 = (2\pi\omega_0^2\chi^{(2)}r_0^2/c^2)A_2$ are normalized complex amplitudes of the fundamental frequency $(\omega = \omega_0)$ and second harmonic $(\omega = 2\omega_0)$ waves; $k_1 = k(\omega_0)$; $k_2 = k(2\omega_0) \approx 2k_1$; $A_{1,2}(\eta,\xi)$ are the slowly varying amplitudes; $r_0$ is the transverse scale of the input beams; $\eta = x/r_0$ is the normalized transverse coordinate; $\xi = z/(k_1 r_0^2)$ is the propagation distance; $\beta = (2k_1 - k_2)k_1 r_0^2$ is the phase mismatch; $d_1 = -1$; $d_2 = -k_1/k_2 \approx -1/2$. Stationary solutions of Eqs. (1) have the form $q_{1,2}(\xi,\eta) = w_{1,2}(\eta)\exp(ib_{1,2}\xi)$, where $w_{1,2}(\eta)$ are real functions, and $b_{1,2}$ are real propagation constants which verify $b_2 = \beta + 2b_1$. The cnoidal wave families are defined by the transverse period $T$ and the propagation constant $b_1$, for a given value of the phase mismatch $\beta$. Physically, $b_1$ is related to the energy flow $U = \int_{-T/2}^{T/2} (w_1^2 + w_2^2)d\eta$ inside each period. Since one can use scaling transformations to obtain periodic waves with different periods from a given family, we selected $r_0$ is such way that the period $T$ equals to $2\pi$, and vary the propagation constant. Particular examples of cnoidal wave solutions of Eqs. (1) can be obtained analytically with the aid of Hamiltonian formalisms, direct substitution, and Lie group techniques [17]. However, whole families of solutions must be obtained numerically [18]. Here we concentrate on two simplest types of cnoidal waves defined by their dn- and cn-type asymptotical shapes in the limit $|\beta| \gg 1$. Examples of the dispersion diagrams and profiles of such dn- and cn-type waves are shown in Figs. 1(a)-1(c) and 2(a)-2(c), respectively. Waves of cn- and dn-types exist for propagation constant values above a cut-off. For cn-waves the cut-off is given by $b_1 = -\beta/2$ at $-\infty < \beta \le 1$, and by $b_1 = -1/2$ at $\beta > 1$. The cut-off for dn-waves is always positive; it monotonically decreases with increase of the phase mismatch, and approaches $b_1 = -\beta/2$ as $\beta \to -\infty$ [18]. In the high-energy, or high-localization limit, cn-type waves transform into arrays of out-of-phase bright solitons, while dn-type waves transform into arrays of in-phase bright solitons. In the low energy limit, dn-type waves transform into plane waves, whereas cn-type waves transform either into plane or small amplitude harmonic waves depending on the phase mismatch sign. Numerical simulations indicate that many of the stationary solutions, including all the solutions obtained analytically [17], destabilize and decay [18], but the stability properties of the entire families is unknown. However, the mathematical techniques commonly used to elucidate the linear stability of localized single soliton solutions (see [16]) are inapplicable to the cnoidal waves. Therefore, here we elaborate a new mathematical formalism put forward to elucidate the stability of such solutions for different degrees of localization. We thus seek for perturbed solutions of Eqs. (1) in the form:

$$q_{1,2}(\eta,\xi) = [w_{1,2}(\eta) + U_{1,2}(\eta,\xi) + iV_{1,2}(\eta,\xi)]\exp(ib_{1,2}\xi), \tag{2}$$

with $w_{1,2}(\eta)$ being the stationary solutions of Eqs. (1), and $U_{1,2}$ and $V_{1,2}$ are respectively real and imaginary parts of small perturbations. We look for exponentially growing perturbations $U_{1,2}(\eta,\xi) = \mathrm{Re}\left[u_{1,2}(\eta,\delta)\exp(\delta\xi)\right]$, $V_{1,2}(\eta,\xi) = \mathrm{Re}\left[v_{1,2}(\eta,\delta)\exp(\delta\xi)\right]$, where $\delta$ is the complex growth rate. Substitution of (2) into (1) and linearization around the stationary solutions yields equations

$$
\begin{gathered}
\frac{d\Phi}{d\eta} = \mathcal{B}\Phi, \quad \mathcal{B} = \begin{pmatrix} \mathcal{O} & \mathcal{E} \\ \mathcal{N} & \mathcal{O} \end{pmatrix}, \\
\mathcal{N} = \begin{pmatrix} -2(b_1 - w_2)/d_1 & 2w_1/d_1 & -2\delta/d_1 & 0 \\ 4w_1/d_2 & -2b_2/d_2 & 0 & -2\delta/d_2 \\ 2\delta/d_1 & 0 & -2(b_1 + w_2)/d_1 & 2w_1/d_1 \\ 0 & 2\delta/d_2 & 4w_1/d_2 & -2b_2/d_2 \end{pmatrix}
\end{gathered}
\tag{3}
$$

for the perturbation vector $\Phi(\eta) = \{u_1, u_2, v_1, v_2, du_1/d\eta, du_2/d\eta, dv_1/d\eta, dv_2/d\eta\}^{\mathrm{T}}$, where $\mathcal{O}$ and $\mathcal{E}$ are zero and unity $4\times4$ matrices, respectively. The general solution of Eqs. (3) can be expressed in the form $\Phi(\eta) = \mathcal{J}(\eta,\eta')\Phi(\eta')$. Here $\mathcal{J}(\eta,\eta')$ is the $8\times8$ Cauchy matrix, that is the solution of the initial value problem $\partial\mathcal{J}(\eta,\eta')/\partial\eta = \mathcal{B}(\eta)\mathcal{J}(\eta,\eta')$, $\mathcal{J}(\eta',\eta') = \mathcal{E}$. The Cauchy matrix defines the matrix of translation of the perturbation eigenvector $\Phi$ on one wave period, as $\mathcal{P}(\eta) = \mathcal{J}(\eta+T,\eta)$. It was rigorously proven in Refs. [13] that the perturbation eigenvector $\Phi_k(\eta)$ is finite along the transverse $\eta$-axis when the corresponding eigenvalue of the matrix of translation satisfies the condition $|\lambda_k| = 1$ $(k = 1,...,8)$. This condition thus defines the algorithm to construct the areas of existence of finite perturbations. The eigenvalues $\lambda_k$ are given by the characteristic polynomial $D(\lambda) = \det(\mathcal{P} - \lambda\mathcal{E}) = \sum_{k=0}^{8} p_k\lambda^{8-k} = 0$, which is independent on $\eta$. The coefficients of the polynomial are given by traces $T_k = \mathrm{Tr}[\mathcal{P}^k(\eta)]$. One finds that $p_0 = p_8 = 1$, $p_1 = p_7 = -T_1$, $p_2 = p_6 = (T_1^2 - T_2)/2$, $p_3 = p_5 = -T_1^3/6 + T_1T_2/2 - T_3/3$, and $p_4 = T_1^4/24 - T_1^2T_2/4 + T_1T_3/3 + T_1^2/8 - T_4/4$. Four of the eight eigenvalues $\lambda_k$ can be omitted because $\lambda_k = 1/\lambda_{k+4}$ $(k = 1,...,4)$, and the corresponding eigenvectors fulfill the symmetry relations: $\Phi_k(\eta) = \Phi_{k+4}(-\eta)$ $(k = 1,...,4)$. The eigenvalues of the translation matrix satisfying $|\lambda_k| = 1$ verify $\lambda_k = \exp(i\mu_kT)$, $-\pi/T \le \mu_k \le \pi/T$, and the Cauchy matrix writes $\mathcal{J}(\eta,\eta') = \mathcal{S}(\eta,\eta')\exp[\mathcal{X}(\eta')(\eta - \eta')]$, where $\mathcal{S}(\eta+T,\eta')=\mathcal{S}(\eta,\eta')$. Thus the eigenvalues $\lambda_k$ of the matrix $\mathcal{P}(\eta)=\exp[\mathcal{X}(\eta)T]$ are connected with the eigenvalues $x_k$ of the matrix $\mathcal{X}(\eta)$ as: $\lambda_k=\exp(x_k T)$, where $x_k=i\mu_k$, and the eigenvectors coincide. Hence, using the Bloch theorem, the solution of Eqs. (3) can be written as $\Phi(\eta)=\mathcal{J}(\eta,\eta')\Phi_k(\eta')=\exp(i\mu_k\eta)\Psi_k(\eta,\eta')$, where $\Psi_k(\eta,\eta')=\Psi_k(\eta+T,\eta')=\mathcal{S}(\eta,\eta')\exp(-i\mu_k\eta')\Phi_k(\eta')$. Our method is advantageous in comparison with standard eigenvalue solvers since only $8\times 8$ matrices are required for construction of areas of existence of finite perturbations. The only time consuming part is the calculation of the translation matrix $\mathcal{P}(\eta)$.

First we apply the method to dn-waves. We searched for all types of perturbations with general complex growth rates, and scanned the whole $\delta$-plane with fine meshes (typically the step in the modulus of $\delta$ was 0.001 and the step in the phase was $\pi/1000$). For periodic waves the areas of existence of finite perturbations have a *band structure*. Typical areas of existence of perturbations with purely imaginary $\delta$ are shown in Fig. 1(d) for $\beta=0$. Such areas overlap with the regions where perturbations exhibit a harmful real growth rate. This is clearly seen in Fig. 1(e) that shows the areas of existence of finite perturbations corresponding to real $\delta$ at $\beta=0$. The corresponding instability is most pronounced at the low-energy limit, when $b_1$ approaches cut-off and the dn-wave transforms into a plane wave, and it is asymptotically suppressed in the high-energy limit, when the dn-wave transforms into an array of in-phase bright solitons. We did not find perturbations with complex $\delta$ (corresponding to oscillatory instabilities) in the case of dn-waves. The typical destabilization of a perturbed dn-wave is shown in Fig. 1(f). The instability of dn-wave always manifests as fusion of neighboring peaks. Similar results were obtained for all values of the propagation constant at different phase-mismatches inside the interval investigated, namely $-20\le\beta\le 20$. Therefore we conclude that the entire families of multicolored dn-waves are unstable.

Next we consider cn-waves. On intuitive grounds, such waves are expected to be more robust than the dn-waves because of the alternating phase of the FF, hence neighboring peaks might tend to repel each other and thus a stable balance might be possible. Still, such is not the case, e.g., in self-focusing cubic Kerr-type nonlinear media where the cn-waves are unstable, too. The central result of this Letter is that such *stable balance does occur with multicolor cn-waves*, as shown in Figs 2(e)-(g). The intrinsically saturable nature of quadratic nonlinearities appears to be important to such stabilization, thus similar results might occur in other systems modeled by saturable nonlinearities. The areas of existence of finite perturbations with imaginary growth rates for cn-waves (Fig. 2(d)) are analogous to that for dn-waves (Fig. 1(d)), but the important result, shown in Figs 2(e)-(g), is that the areas of existence of perturbations with growth rates having nonzero real part *vanish* when the propagation constant *exceeds a certain critical value* $b_{\rm cr}$. Therefore, above the corresponding energy threshold, the families of multicolor cn-waves become *completely* free of linear instabilities with exponential growth. Figures 2(e) and 2(f) indicate that the threshold energy for stabilization decreases when the phase mismatch $\beta$ goes from $-\infty$ to approximately 0.25. Thus, stable cn-waves occur almost in the whole range of its existence at small positive and negative $\beta$. For example, at $\beta = -3$ we have found stable cnoidal waves for $b_1 \geq b_{\rm cr} \approx 2.382$, while at $\beta = 0$ the region of stability begins at $b_{\rm cr} \approx 0.192$. The exponential instabilities of low-energy cn-waves associated to purely real growth rates exist at $-\infty < \beta \leq 1$. In the region $0.25 \lesssim \beta < \infty$ we have found oscillatory instabilities associated to complex growth rates. Such instabilities also cease to exist when the energy flow exceeds a threshold value. Fig. 2(g) illustrates this point for $\beta = 3$. For oscillatory instabilities the critical value of the propagation constant also decreases as $\beta$ goes from $\infty$ to approximately 0.25. At fixed $\beta$ and energy flow, the complex growth rates form the curves in the complex plane $[\mathrm{Re}(\delta), \mathrm{Im}(\delta)]$ shown in Fig. 2(h). The key result of this paper is thus summarized in Fig. 2(i), which shows the threshold value of the propagation constant for complete stabilization versus phase mismatch. The left curve stands for exponential instabilities while the right curve stands for oscillatory instabilities.

To confirm the predictions of the linear stability analysis, we performed series of numerical simulations of Eqs. (1) with the input conditions $q_{1,2}(\eta, \xi = 0) = w_{1,2}(\eta)[1 + \rho_{1,2}(\eta)]G(\eta)$, where $w_{1,2}(\eta)$ are profiles of the stationary waves, $\rho_{1,2}(\eta)$ is a random function with a Gaussian distribution and variance $\sigma_{1,2}^2$, and the function $G(\eta)$ is a broad Gaussian envelope imposed on the periodic cnoidal pattern. The width of the envelope was much larger than the cnoidal wave period. Figs 3(a)-(d) illustrate the outcome of the numerical simulations. The typical decay of low-energy cn-waves is shown in Fig. 3(a). Beyond the point of spontaneous onset of the instability, the wave is rapidly destroyed. In clear contrast, Figs 3(b)-(d) illustrate the stable propagation of perturbed cnoidal waves for several values of phase mismatch and energy flow. The cnoidal waves kept their input structure for all the monitored propagation distances, in some cases over several thousand units.

Therefore, in conclusion, the stability analysis of periodic cnoidal wave arrays in quadratic medium performed by a newly developed mathematical formalism has revealed the existence of completely stable patterns in the form of cn-waves. From an applied point of view, such patterns of periodic pixel-like structures can find applications, e.g., in the study of complex light patterns generated by modulational instabilities, or in the implementation of digital image processing schemes based on soliton-like arrays [19]. From a fundamental viewpoint, we stress that the reported multicolor cn-waves are the first known example in physics of stable periodic waves propagating in a conservative uniform medium supporting bright solitons.

## Figure captions

Figure 1. (a) Energy flow of dn-waves versus propagation constant for various phase mismatches. (b), (c) dn-wave profiles at $\beta = 0$ and $b_1 = 0.6$ and $b_1 = 1.5$, respectively. (d) Areas of existence of finite perturbations with imaginary growth rates at $\beta = 0$ (shaded). (e) Areas of existence of finite perturbations with real growth rates at $\beta = 0$ (shaded). Cut-off on propagation constant in figures (d) and (e) is given by $b_1 \approx 0.341$. (f) Propagation of dn-wave with $b_1 = 0.6$ at $\beta = 0$ in the presence of the perturbation with $\delta = 0.12444$. Only the FF wave is shown.

Figure 2. (a) Energy flow of cn-waves versus propagation constant for various phase mismatches. (b), (c) cn-wave profiles at $\beta = 0$ and $b_1 = 0.15$ and $b_1 = 1.5$, respectively. (d) Areas of existence of finite perturbations with imaginary growth rates at $\beta = 0$ (shaded). (e), (f) Areas of existence of finite perturbations with real growth rates (shaded) at $\beta = -3$ and $\beta = 0$, respectively. (g) Maximum real part of complex growth rate versus propagation constant at $\beta = 3$. (h) Curves at the complex plane showing possible increment values for various propagation constants at $\beta = 3$. (i) Threshold propagation constant for stabilization versus phase mismatch.

Figure 3. (a) Propagation of the unstable cn-wave with $b_1 = 0.15$ at $\beta = 0$ in the presence of the perturbation with $\delta = 0.0954$. Long-distance evolution of stable cn-waves with $b_1 = 1.8$ at $\beta = -2$ (b), $b_1 = 0.5$ at $\beta = 0$ (c), $b_1 = 0.5$ at $\beta = 1$ (d) in the presence of white noise with variance $\sigma_{1,2}^2 = 0.01$. Only the FF wave is shown.

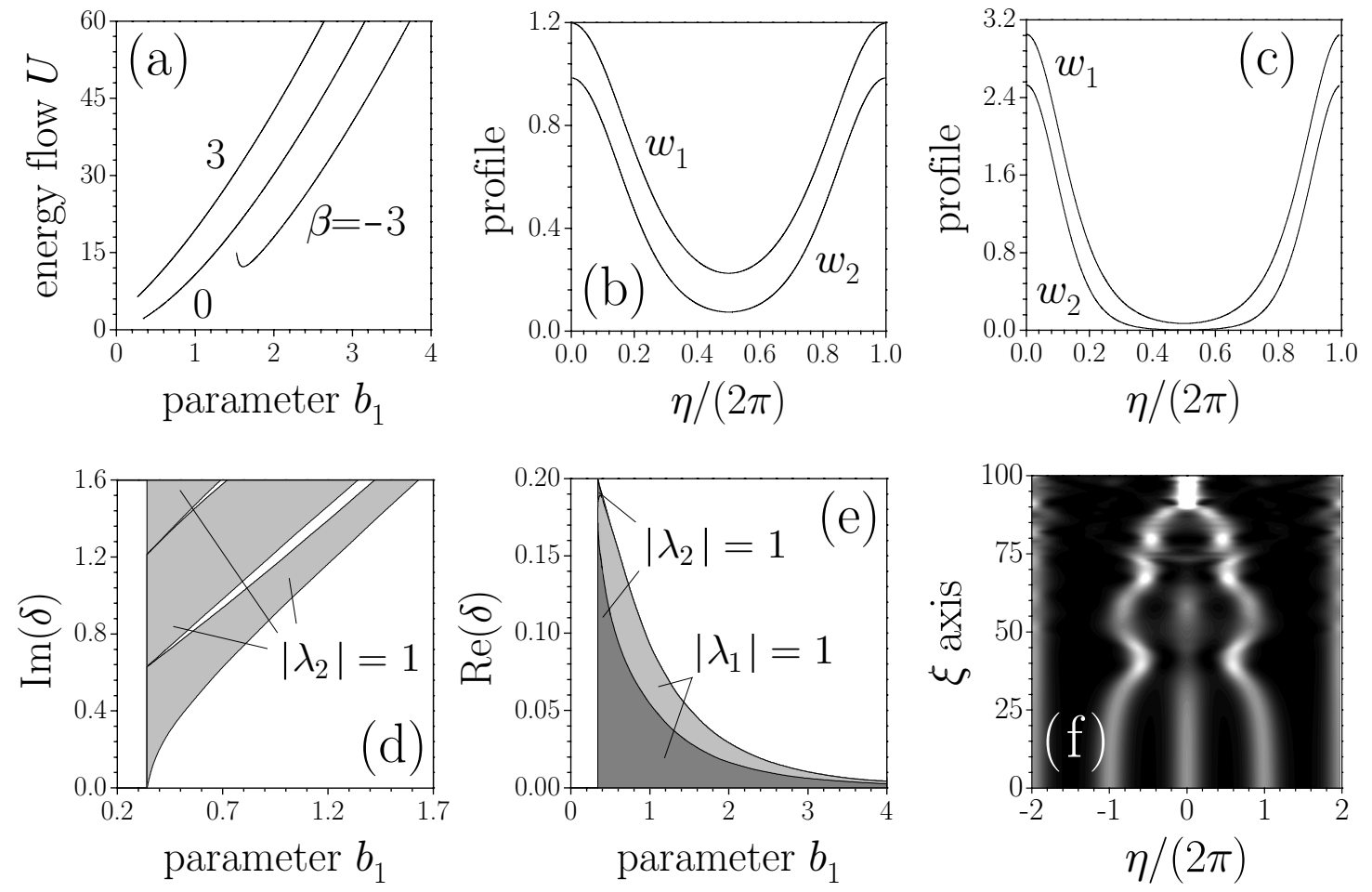

Figure 1. (a) Energy flow of dn-waves versus propagation constant for various phase mismatches. (b), (c) dn-wave profiles at $\beta = 0$ and $b_1 = 0.6$ and $b_1 = 1.5$, respectively. (d) Areas of existence of finite perturbations with imaginary growth rates at $\beta = 0$ (shaded). (e) Areas of existence of finite perturbations with real growth rates at $\beta = 0$ (shaded). Cut-off on propagation constant in figures (d) and (e) is given by $b_1 \approx 0.341$. (f) Propagation of dn-wave with $b_1 = 0.6$ at $\beta = 0$ in the presence of the perturbation with $\delta = 0.12444$. Only the FF wave is shown.

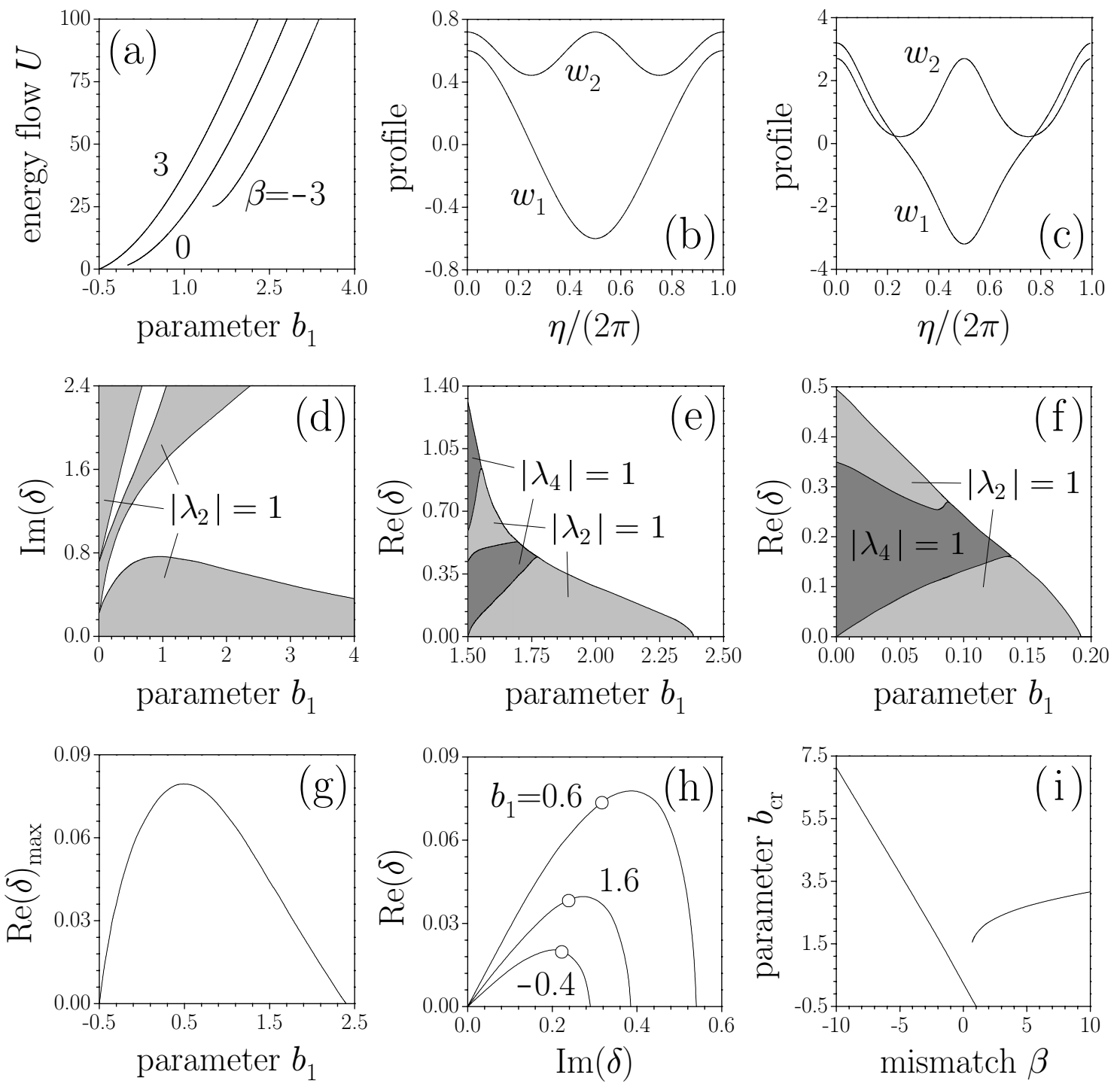

Figure 2. (a) Energy flow of cn-waves versus propagation constant for various phase mismatches. (b), (c) cn-wave profiles at $\beta = 0$ and $b_1 = 0.15$ and $b_1 = 1.5$, respectively. (d) Areas of existence of finite perturbations with imaginary growth rates at $\beta = 0$ (shaded). (e), (f) Areas of existence of finite perturbations with real growth rates (shaded) at $\beta = -3$ and $\beta = 0$, respectively. (g) Maximum real part of complex growth rate versus propagation constant at $\beta = 3$. (h) Curves at the complex plane showing possible increment values for various propagation constants at $\beta = 3$. (i) Threshold propagation constant for stabilization versus phase mismatch.

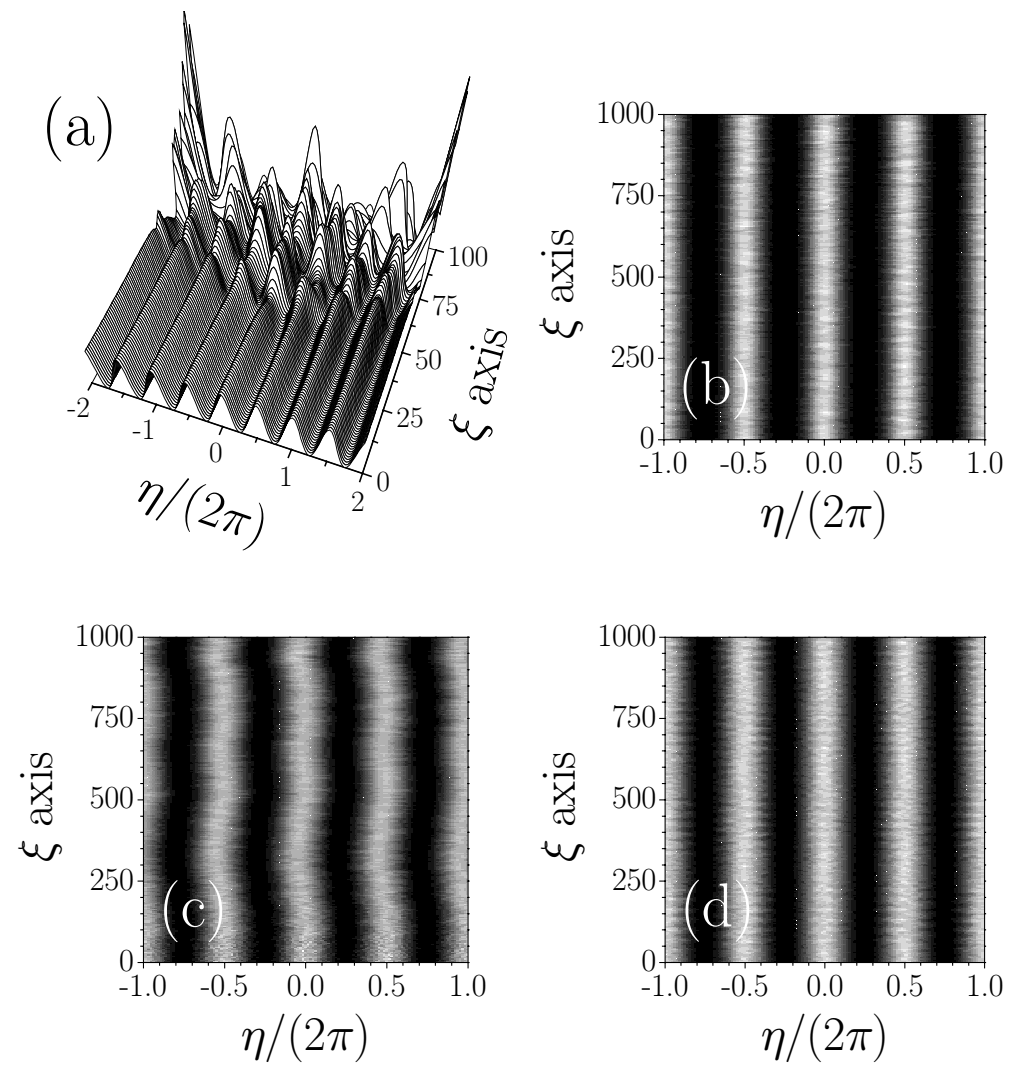

Figure 3. (a) Propagation of the unstable cn-wave with $b_1 = 0.15$ at $\beta = 0$ in the presence of the perturbation with $\delta = 0.0954$. Long-distance evolution of stable cn-waves with $b_1 = 1.8$ at $\beta = -2$ (b), $b_1 = 0.5$ at $\beta = 0$ (c), $b_1 = 0.5$ at $\beta = 1$ (d) in the presence of white noise with variance $\sigma_{1,2}^2 = 0.01$. Only the FF wave is shown.